\documentclass[preprint,superscriptaddress,amsmath,amssymb,prd,aps,showpacs,floatfix,nofootinbib]{revtex4-1}
\usepackage{graphicx}
\usepackage{dcolumn}
\usepackage{bm}
\usepackage{mathrsfs}
\usepackage{hyperref}
\usepackage{amsmath}
\usepackage{amssymb}
\usepackage{float}
\usepackage{dcolumn}
\usepackage{multirow}

\begin{document}
\title{Analysis of the $\psi(4040)$ and $\psi(4160)$ decay into $D^{(\ast)} \bar D^{(\ast)}$,  $D_s^{(\ast)}\bar D_s^{(\ast)}$}
\date{\today}
\author{M.~Bayar}
\email{melahat.bayar@kocaeli.edu.tr}
\affiliation{Department of Physics, Kocaeli University, Izmit 41380, Turkey}
\affiliation{Departamento de
F\'{\i}sica Te\'orica and IFIC, Centro Mixto Universidad de
Valencia-CSIC Institutos de Investigaci\'on de Paterna, Aptdo.22085,
46071 Valencia, Spain}

\author{N.~Ikeno}
\email{ikeno@tottori-u.ac.jp}
\affiliation{Department of Agricultural, Life and Environmental Sciences, Tottori University, Tottori 680-8551, Japan}
\affiliation{Departamento de
F\'{\i}sica Te\'orica and IFIC, Centro Mixto Universidad de
Valencia-CSIC Institutos de Investigaci\'on de Paterna, Aptdo.22085,
46071 Valencia, Spain}

\author{E.~Oset}
\email{oset@ific.uv.es}
\affiliation{Departamento de
F\'{\i}sica Te\'orica and IFIC, Centro Mixto Universidad de
Valencia-CSIC Institutos de Investigaci\'on de Paterna, Aptdo.22085,
46071 Valencia, Spain}
\affiliation{Department of Physics, Guangxi Normal University, Guilin 541004, China}

\begin{abstract}
We have performed an analysis of the $e^+ e^- \to D^{(*)} \bar D^{(*)}$ data in the region of the  $\psi(4040)$ and $\psi(4160)$ resonances which have a substantial overlap and require special care. By using the $^3 P_0$ model to relate the different $D^{(*)} \bar D^{(*)}$ production modes, we make predictions for production of these channels and compare with experiment and other theoretical approaches. As a side effect we find that these resonances qualify largely as $c \bar c$ states and the weight of the meson-meson components in the wave function is very small.  
\end{abstract}

\maketitle


\section{Introduction}

The charmonium spectroscopy has been studied both theoretically and experimentally over four decades, but there are still many debates in charmonium  physics. Although the charmonium system is well described below the open charm threshold, for the charmonium states above the open charm threshold  there is no real consensus between experimental information and theoretical results. Above the open charm threshold charmonium states can decay into several two body final states  as stated in Refs. \cite{Eichten:1979ms, Barnes:2005pb}, the decay channels are $\psi(4040) \to D\bar D, D\bar D^\ast, D^\ast\bar D, D^\ast\bar D^\ast, D_{s}\bar D_{s}$, $\psi(4160) \to D\bar D, D\bar D^\ast, D^\ast\bar D, D^\ast\bar D^\ast, D_{s}\bar D_{s},  D_{s}\bar D^{*}_{s},  D^{*}_{s}\bar D_{s}$. It is also important to consider charm meson production to understand the hadron spectrum well.

Experimental research has been done to determine masses and total widths  of the $\psi(4040)$ and $\psi(4160)$  states. For instance, the total cross section for hadron production in $e^{+}e^{-}$ annihilation was analysed in Ref.~\cite{Ablikim:2007gd}. They measured $R$ values, which is defined as $R=\sigma (e^{+}e^{-} \to ~hadrons)/ \sigma (e^{+}e^{-} \to ~\mu^{+} \mu^{-}) $  with the BESII detector at center-of-mass energies between $3.7$ and $5.0$ GeV. They show the $R$ values for the high mass charmonia structure in Fig.~1 of Ref.~\cite{Ablikim:2007gd} where the two resonances have some overlap. In the analysis they use an energy dependent total width and introduce relative phases between the resonances in the fit. They conclude that  the results are sensitive to the form of the energy-dependent total width, and the influence of the phase angles of the resonance parameters is important.  The $e^{+}e^{-}$ annihilation with exclusive production of $ D\bar D, D\bar D^\ast, D^\ast\bar D^\ast$ was investigated in Ref.  \cite{Aubert:2009aq}. The authors mention that interference between the resonances and noresonance contributions is required to obtain a satisfactory description of the data, 
somewhat a different interpretation than that of Ref.~\cite{Ablikim:2007gd}. 
They also measured $\Gamma(\psi(4040)\to D\bar D)/ \Gamma(\psi(4040)\to D^{*} \bar D) $ and find $0.24\pm 0.05 \pm 0.12$ and compare with a calculation using the $^{3}P_{0}$ model of Ref. \cite{Barnes:2005pb}. The agreement is not good. It is not clear how the spin-angular momentum algebra is done for the $^{3}P_{0}$ model in Ref. \cite{Barnes:2005pb}. 

The quantum numbers assigned to the $\psi(4040)$ and $\psi(4160)$ are $ 3^{3}S_{1}$ and  $ 2^{3}D_{1}$ in Ref.~\cite{Barnes:2005pb}, respectively. The same assignment is done in Ref.~\cite{Godfrey:1985xj,Segovia:2013wma,Ortega:2017qmg}. The decay widths of the $\psi(4040)$  and $\psi(4160)$ had been evaluated in the $^{3}P_{0}$ model (Table X and XII in Ref.~\cite{Barnes:2005pb}) and also  in Ref.~\cite{Aubert:2009aq} (Table VI). The results  are very different.  In Ref.~\cite{Goldhaber:1977qn}, the authors calculate  $\Gamma(\psi(4040)\to D^{0}\bar D^{0})/ \Gamma(\psi(4040)\to D^{0*} \bar D^{0}+~cc) =0.05\pm 0.03$. Yet, this seems to be in contradiction with the results of Ref.~\cite{Aubert:2009aq}.

As one can see, there is no noconsensus on the interpretation of the data. Inclusive and exclusive cross sections for $D^+$, $D^-$, and $D_s^{*}$ production in $e^+ e^-$ annihilation are also available from Ref.~\cite{CroninHennessy:2008yi}. 
There are further data from Belle~\cite{Zhukova:2017pen}, where measurements of $e^+ e^- \to D^{(*) \pm} D^{* \mp}$ using initial state radiation are done. Further data from BES on $e^+ e^- \to hadrons$ are available in~\cite{Bai:1999pk,Bai:2001ct}, which are analyzed in~\cite{Mo:2010bw} in terms of three resonances, $\psi(4040), \psi(4160), \psi(4415)$, with interference among them. We shall not consider the $\psi(4415)$, which is far away and disconnected from the other two resonances and concentrate on the $\psi(4040)$ and $\psi(4160)$ which overlap in the hadron production spectrum.

This paper is organized as follows. In Sec.~\ref{formalism}, we establish the formalism of calculating the cross section for $e^+e^-\to $ hadrons through the dressed propagator of the $\psi(4040)$  and $\psi(4160)$, In Sec.~\ref{result}, we present the results on the line shape of the $\psi(4040)$  and $\psi(4160)$ fitting to the experimental data, the meson-meson probabilities in the $\psi(4040)$  and $\psi(4160)$ wave functions and the $Z$ probabilities using the parameters extracted from the fitting. A summary and conclusion is presented in Sec.~\ref{conclusion}.

\section{Formalism}\label{formalism}
\begin{figure}[h!]
  \centering
  \includegraphics[width=0.90\textwidth]{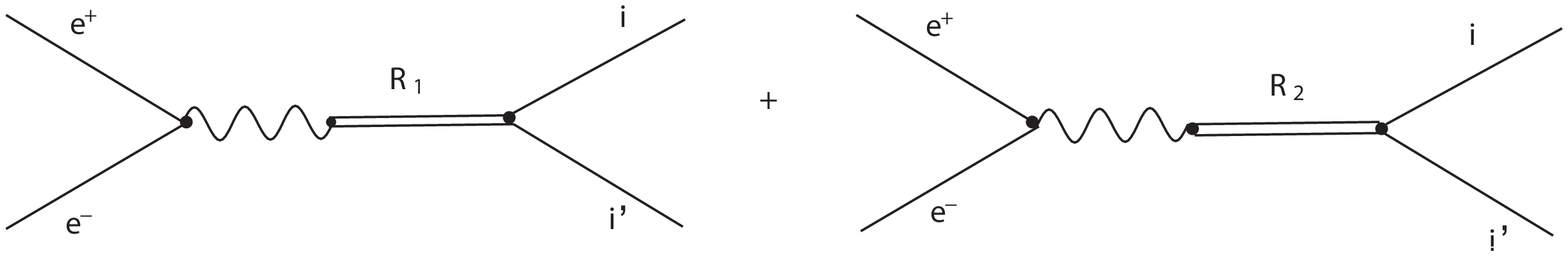}
  \caption{Feynman diagrams of $e^{+}e^{-} \to R_{1} ( R_{2}) \to i~i' $, where $R_{1}\equiv \psi(4040) $ and $R_{2}\equiv \psi(4160)$ and $i, i'$ correspond to any of the $D^{0}\bar D^{0}$, $D^{+}\bar D^{-}$, $D^{0}\bar D^{*0}$, $D^{*0}\bar D^{0}$, $D^{*+}\bar D^{-}$, $D^{+}\bar D^{*-}$, $D^{*0}\bar D^{*0}$, $D^{*+}\bar D^{*-}$, $D_{s}^{+}\bar D_{s}^{-}$, $D_{s}^{+}\bar D_{s}^{*-}$, $D_{s}^{*+}\bar D_{s}^{-}$, $D_{s}^{*+}\bar D_{s}^{*-}$ channels.}
  \label{epem}
\end{figure}

\begin{figure}[h!]
  \centering
  \includegraphics[width=0.60\textwidth]{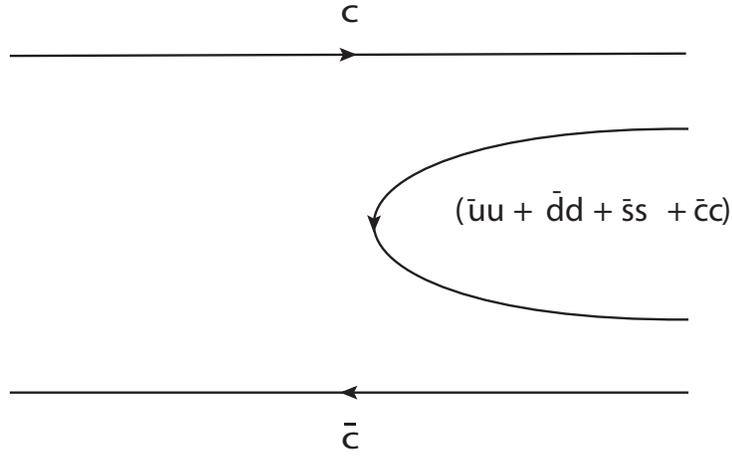}
  \caption{Hadronization process for $\psi(4040) [\psi(4160)] \to i i'$.}
  \label{had}
\end{figure}

The Feynman diagrams of hadron production in $e^+ e^-$ annihilation for  $\psi(4040)$ and $\psi(4160)$ are shown in Fig.~\ref{epem}. The first part of the Feynman diagrams in Fig.~\ref{epem} is the electromagnetic interaction and the second part of the diagrams is strong interaction.
In the second part, we assume the $\psi$ resonances to be $c \bar{c}$ states and insert  a $\bar qq$ pair with the quantum numbers of the vacuum  between the $c \bar{c}$, the quark components of the $\psi(4040)$ and $\psi(4160)$, in order to produce the $D^{(*)} \bar{D}^{(*)}$ pairs. This is called hadronization and the mechanism of hadronization  is shown in Fig.~\ref{had}. 

We show the hadronization procedure in detail. We will include a $ \bar q q$ pair  with the  quantum  numbers  of  the  vacuum and will obtain a pair of mesons. These meson pairs could be vector-vector ($VV$), pseudoscalar-vector ($PV$), vector-pseudoscalar ($VP$) or pseudoscalar-pseudoscalar ($PP$) meson. To do this, we first write the $ \bar q q$ matrix $M$

\begin{equation}\label{matrix1}
M = (q\bar q)=\begin{pmatrix}
u\bar u & u\bar d & u\bar s & u\bar c \\
d\bar u & d\bar d & d\bar s & d\bar c \\
s\bar u & s\bar d & s\bar s & s\bar c \\
c\bar u & c\bar d & c\bar s & c\bar c \\
\end{pmatrix}.
\end{equation}
The hadronization in the flavor space proceeds as 
\begin{equation}
 c \bar c \to  c \sum_{i=1}^{4} \ \bar q_{i} q_{i}\ \bar c =  \sum_{i=1}^{4} \ c \bar q_{i} q_{i}  \bar c = (M M)_{44}.
\end{equation}

On the hadron level we can write the $M$ matrix in terms of the physical mesons. This  matrix $M$ could be the vector or the pseudoscalar matrix. The vector matrix $V$ is given by 

\begin{equation}\label{Vecmatrix}
V = \begin{pmatrix}
 \frac{1}{\sqrt{2}}\rho^0 + \frac{1}{\sqrt{2}} \omega & \rho^+ & K^{* +} & \bar{D}^{* 0} \\
 \rho^- & -\frac{1}{\sqrt{2}}\rho^0 + \frac{1}{\sqrt{2}} \omega  & K^{* 0} & \bar{D}^{* -} \\
 K^{* -} & \bar{K}^{* 0}  & \phi & D_s^{* -} \\
 D^{* 0} & D^{* +} & D_s^{* +} & J/\psi
\end{pmatrix}.
\end{equation}
and the pseudoscalar matrix can be written as 

\begin{equation}\label{Psematrix}
\phi = \begin{pmatrix}
\frac{1}{\sqrt{2}}\pi^0 + \frac{1}{\sqrt{3}} \eta + \frac{1}{\sqrt{6}}\eta' & \pi^+ & K^+ & \bar{D}^0 \\
 \pi^- & -\frac{1}{\sqrt{2}}\pi^0 + \frac{1}{\sqrt{3}} \eta + \frac{1}{\sqrt{6}}\eta' & K^0 & D^- \\
 K^- & \bar{K}^0 & -\frac{1}{\sqrt{3}} \eta + \sqrt{\frac{2}{3}}\eta' & D_s^- \\
D^0  & D^+ & D_s^+ & \eta_c
\end{pmatrix}.
\end{equation}
where the standard  $ \eta - \eta'$ mixing of Ref.~\cite{Bramon:1992kr} is used. Finally  as we mentioned before, the hadronization gives rise to  $PP$, $PV$, $VP$, $VV$ in terms of two mesons and hence we have 
\begin{align}
(\phi\phi)_{4,4}&=D^0\bar D^0+D^+D^-+D_s^+D_s^-, \label{eq:phi.phi} \\ 
(PV)_{4,4}&=D^0\bar D^{\ast 0}+D^+D^{\ast-}+D_s^+D_s^{\ast-},\\
(VP)_{4,4}&=D^{\ast 0}\bar D^0+D^{\ast+}D^-+D_s^{\ast+}D_s^-,\\
(VV)_{4,4}&=D^{\ast0}\bar D^{\ast 0}+D^{\ast+}D^{\ast-}+D_s^{\ast+}D_s^{\ast-}. \label{eq:VV}
\end{align}
where we have neglected $\eta_c^2$,  $\eta_c \ J/\psi$, $J/\psi \ J/\psi$, since they are too heavy relative to the other channels.

Let us come back again to Fig.~\ref{epem}. First there is  an interaction  $e^+ e^-\to R_{1(2)}$ and then decay of $R_{1(2)}\to i~i'$ where $i~i'$ correspond to the $D^{0}\bar D^{0}$, $D^{+}\bar D^{-}$, $D^{0}\bar D^{*0}$, $D^{*0}\bar D^{0}$, $D^{*+}\bar D^{-}$, $D^{+}\bar D^{*-}$, $D^{*0}\bar D^{*0}$, $D^{*+}\bar D^{*-}$, $D_{s}^{+}\bar D_{s}^{-}$, $D_{s}^{+}\bar D_{s}^{*-}$, $D_{s}^{*+}\bar D_{s}^{-}$, $D_{s}^{*+}\bar D_{s}^{*-}$ channels. The couplings of  $e^+ e^-\to R_{1}$ and  $e^+ e^-\to R_{2}$ are different and  the couplings of  $R_{1}\to i~i'$ and  $R_{2}\to i~i'$  are also different. However, for each resonance we can relate the amplitude for $D^{0}\bar D^{0}$, $D^{+}\bar D^{-}$, $D^{0}\bar D^{*0}$, $D^{*0}\bar D^{0}$, $D^{*+}\bar D^{-}$, $D^{+}\bar D^{*-}$, $D^{*0}\bar D^{*0}$, $D^{*+}\bar D^{*-}$, $D_{s}^{+}\bar D_{s}^{-}$, $D_{s}^{+}\bar D_{s}^{*-}$, $D_{s}^{*+}\bar D_{s}^{-}$, $D_{s}^{*+}\bar D_{s}^{*-}$ production by means of the $^{3}P_0$ model used in the paper of Ref. \cite{biz}. 
In this model one uses the most basic $\bar q q$ configuration that provides the $\bar q q$ state with vacuum quantum numbers. Since $\bar q$ has negative parity we need $L=1$ to restore the parity, which implies that the spin $S$ is $S=1$ such that the total angular momentum $J$ can be $J=0$. The spin angular momentum components are considered together with the spin of the $c \bar c$ components to finally provide the spin of the $D^{(*)} \bar D^{(*)}$ pairs, which are produced in $p$-wave. Details on the angular momentum algebra for the process are given in Ref.~\cite{biz}.

Now let us write the matrix element of the first part (electron-positron pair annihilation) of the Feynman diagrams of Fig.~\ref{epem} as 
\begin{equation}\label{matrixel1}
t'\sim e \bar{v}(p_{1}) \gamma^{\mu} u(p_{2}) \epsilon_{\mu}(\gamma)
\end{equation}
where $u(p_{2})$ and $\bar{v}(p_{1})$ are the spinors of the electron and positron,  respectively and $\epsilon_{\mu}(\gamma)$ is the polarization of the photon. Then the square of the absolute value of the matrix element is obtained as follows
\begin{equation}\label{matrixel2}
|t'|^{2}\sim \dfrac{4 e^{2}}{2 m 2 m} (p^{\mu}_{1} p^{\nu}_{2} -p_{1}.p_{2} g^{\mu \nu}+ p^{\nu}_{1} p^{\mu}_{2}) \epsilon_{\mu} \epsilon_{\nu}
\end{equation}
where the original photon polarizations $ \epsilon_{\mu} \epsilon_{\nu}$ become the $R$ polarization after $\gamma \to \psi$ conversion, and in the $R$ rest frame only the spatial componemts of $\epsilon_{\mu}$ remain. 
Then we write the whole matrix element as 
\begin{equation}\label{matrixelt1}
t\sim e \bar{v}(p_{1}) \gamma^{\mu} u(p_{2}) \epsilon_{\mu} f_{R} ~D_{R} ~A_{Ri} ~p'~ \frac{1}{p^{2}_{\gamma}}
\end{equation}
where $D_{R}$ is a propagator of the R meson,  $f_{R}$ is a $\gamma \to \psi$ conversion factor and $A_{Ri}$ the coupling of the R to each $i~i'$ channels ($D^{0}\bar D^{0}$, $D^{+}\bar D^{-}$, $D^{0}\bar D^{*0}$, $D^{*0}\bar D^{0}$, $D^{*+}\bar D^{-}$, $D^{+}\bar D^{*-}$, $D^{*0}\bar D^{*0}$, $D^{*+}\bar D^{*-}$, $D_{s}^{+}\bar D_{s}^{-}$, $D_{s}^{+}\bar D_{s}^{*-}$, $D_{s}^{*+}\bar D_{s}^{-}$, $D_{s}^{*+}\bar D_{s}^{*-}$). We can take the spatial component of $\epsilon_{\mu}$ contracted in $|t|^2$ as $\sum \epsilon_{i}  \epsilon_{j} \to \delta_{ij} $ with $i,j=1,2,3$, so we have 

\begin{equation}\label{matrixelt2}
|t|^{2}\sim \dfrac{e^{2}}{2 m 2 m} M^{2}_{\rm inv} \vert f_{R} ~D_{R} ~A_{Ri} ~p'\vert^{2} \left(  \frac{1}{p^{2}_{\gamma}} \right)^{2}.
\end{equation}
Once we get $|t|^{2}$, we can calculate the cross section as follows
\begin{equation}\label{crosssection}
\sigma = 4 \pi \dfrac{d\sigma}{d\Omega}\sim \dfrac{1}{16 ~\pi} \frac{1}{s} e^{2} M^{2}_{\rm inv} \left(  \frac{1}{p^{2}_{\gamma}} \right)^{2}  \frac{p'}{p}  \vert f_{R} ~D_{R} ~A_{Ri} ~p'\vert^{2}.
\end{equation} 
Using $ M_{\rm inv} = \sqrt{s} $, then the cross section read as 
\begin{equation}\label{crosssectionS}
\sigma =  \dfrac{1}{8 ~\pi} \frac{1}{s^{2}} \frac{1}{\sqrt{s}} e^{2}  (p')^{3}  \vert f_{R} ~D_{R} ~A_{Ri}\vert^{2},
\end{equation}
where $p'$ is the meson momentum in the $M_{i} M'_{i}$ rest frame 

\begin{equation}\label{pppr}
p'=\dfrac{\lambda^{1/2} (s, M_{i}^{2}, M_{i}'^{2})}{2 \sqrt{s}},
\end{equation}
with $M_{i}$ and $M'_{i}$ the masses of the $i, i'$ mesons and $D_{R}$ the propagator of the $R$ vector,
\begin{equation}\label{pppr}
D_{R}=\dfrac{1}{p^{2}-M_{R}^{2}+i M_{R} \Gamma_{R}},
\end{equation}
with $p=p_{1}+p_{2}$, in the case of a free vector. 

\begin{figure}[h!]
  \centering
  \includegraphics[width=0.60\textwidth]{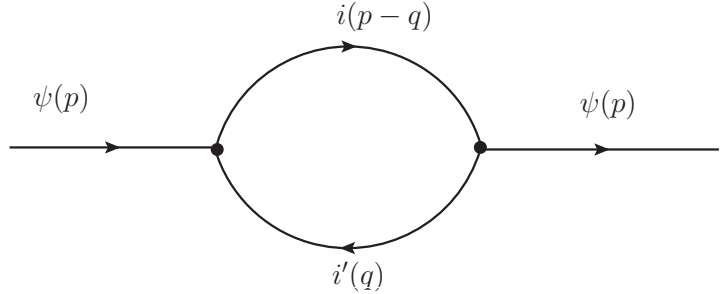}
  \caption{The $\psi$ propagator dressed with a $i~i'$ loop.}
  \label{loop}
\end{figure}

Next, we include the meson-meson self energy in the vector propagator 
\begin{equation}\label{DRP}
D_{R}=\dfrac{1}{p^{2}-M_{R}^{2}-\Pi(p)}.  
\end{equation}

The $\psi$ selfenergy diagrams are depicted in Fig.~\ref{loop}. In order to evaluate $\Pi(p^2)$ we indicate the dressing of $\psi$ by meson-meson components and we include $PP$, $VP$, $PV$, and $VV$. There is only one unknown global parameter $A_{Ri}$ since all the different $D^{(*)} \bar D^{(*)}$, $D^{(*)}_s \bar D^{*}_s$ channels can be connected via Eqs.~(\ref{eq:phi.phi})-(\ref{eq:VV}) and the $^3 P_0$ model, for which the details are given in~\cite{biz}. In that work a detailed derivation was done for the case where the $c \bar c$ quarks in the $\psi$ state are in $D$-wave, which is case for the $\psi(4160)$ according to Ref.~\cite{Godfrey:1985xj} ($2 ^3 D_1$). For $S$-wave, where the $\psi(4040)$ is classified according to Ref.~\cite{Godfrey:1985xj} ($3 ^3 S_1$), the calculations are done in a similar way and we provide here the new coefficients in analogy to those obtained for $D$-wave in~\cite{biz}, where the $\psi(3770)$ ($1 ^3 D_1$ according to Ref.~\cite{Godfrey:1985xj}) was studied.

\begin{figure}[tb!]
  \centering
  \includegraphics[width=1.00\textwidth]{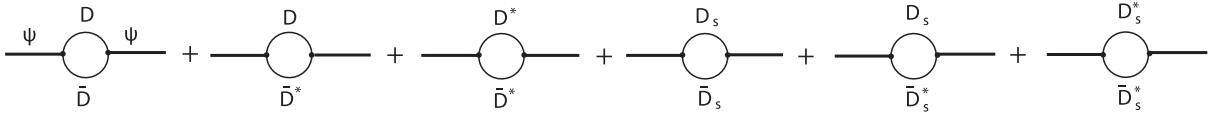}
  \caption{Contribution to the $\psi$ selfenergy for the vector $\psi$ propagator dressed with a meson-meson loop with $cc$.}
  \label{selfEloops}
\end{figure}

The self energy contributions $\Pi (p)$ are shown in Fig.~\ref{loop}, and detailing the channels in Fig.~\ref{selfEloops}, and is written as follows 
\begin{align}
-i\Pi_i (p)=\int\frac{d^4q}{(2\pi)^4}(-i)V_1(-i)V_2\frac{i}{q^2-m^2_{D_{i}}+i\varepsilon}\frac{i}{(p-q)^2-m^2_{D_{i'}}+i\varepsilon}F(\bm q)^2,
\end{align}
where $p$ is the total four momentum of the system, and $m_{D_{i}}$ and $m_{D_{i'}}$ are the masses of the mesons in the $i,i'$-channel.
$F(\bm q)$ is a form factor and the subindex of the $D$ mesons indicates all twelve channels which means  $i, i' =1,2, .~.~,12$. We use the couplings of $\psi(4040)$ or $~\psi(4160) \to PP$, $PV$, $VP$, $VV$ which have the effective form evaluated in the $^3 P_0$ model, such that the sum over polarizations on the vector states in $\Pi$, Fig.~\ref{loop}, and in the final states, Fig.~\ref{epem}, are already done,
\begin{align}
V_{\psi,(D_{i}D_{i'})} = A_{Ri}\ |\bm q|  = A \ g_{R,i} \ |\bm q| ,
\label{eq:19}
\end{align}
where we separate the coupling $A_{Ri}$ of $R \to$ meson-meson channel into a global constant $A$ and the coefficient $g_{R,i}$ which stems from the $^3 P_0$ model calculation.
Then $\Pi(p)$ is rewritten as
\begin{align}
\Pi_i(p)=i\,g^2_{R,i} A^2 \int\frac{d^4q}{(2\pi)^4}\bm q^2\frac{1}{q^2-m^2_{D_{i}}+i\varepsilon}\ \frac{1}{(p-q)^2-m^2_{D_{i'}}+i\varepsilon}F(\bm q)^2.
\end{align}

The $q^0$ integration can be evaluated analytically and then we get in the rest frame of the $\psi(4040)$  and $\psi(4160)$ $(p^0=\sqrt s)$
\begin{align}\label{SelfEn}
\Pi(p^0)=  A^{2} \sum g^2_{R,i}\ \tilde G_i (p^0),
\end{align}
where $A$ is a global constant to be determined from the data, and $\tilde G_i(p^0)$ is 
\begin{align}\label{Gtilda}
\tilde G_i (p^0)=\int\frac{dq}{(2\pi)^2}\frac{\omega_1(\bm q)+\omega_2(\bm q)}{\omega_1(\bm q)\omega_2(\bm q)}\frac{\bm q^4}{(p^0)^2-(\omega_1(\bm q)+\omega_2(\bm q))^2+i\varepsilon}F(\bm q)^2,
\end{align}
where $\omega_1(\bm q)=\sqrt{\bm q^2+m^2_{D_{i}}}$, $\omega_2(\bm q)=\sqrt{\bm q^2+m^2_{D_{i'}}}$. Then the $ g^2_{R,i}$ values in Eq.~(\ref{SelfEn}) are calculated for the $\psi(4040)$ ( $\psi(4160)$) like in Ref.~\cite{biz} and the values are presented in Table~\ref{gvalues}.

 \begin{table*}[tbh!]
\renewcommand\arraystretch{0.5}
\centering
\caption{\vadjust{\vspace{-2pt}} $g^2_{R,i}$ values  for $\psi(4040)$ and $\psi(4160)$ vector mesons.}\label{gvalues}
\begin{tabular*}{1.00\textwidth}{@{\extracolsep{\fill}} c  c c}
\hline
\hline
  Channels    & $g^2_{R,i}$ for $\psi(4040)$  & $g^2_{R,i}$ for  $\psi(4160)$   \\
\hline 
 $D^{0}\bar D^{0}$  & $1/12$  & $1/12$ \\

 $D^{+}\bar D^{-}$& $1/12$  & $1/12$ \\

$D^{0}\bar D^{*0}$ & $1/6$  & $1/24$ \\

$D^{*0}\bar D^{0}$ &  $1/6$  & $1/24$ \\

 $D^{+}\bar D^{*-}$ &  $1/6$  &$1/24$ \\

$D^{*+}\bar D^{-}$ &   $1/6$ &   $1/24$ \\
  
$D^{*0}\bar D^{*0}$ & $7/12$  & $77/120$\\
 
 $D^{*+}\bar D^{*-}$ & $7/12$  & $77/120$ \\

  $D_{s}^{+}\bar D_{s}^{-}$ & $1/12$  &$1/12$ \\
 
 $D_{s}^{+}\bar D_{s}^{*-}$& $1/6$   & $1/24$ \\

 $D_{s}^{*+}\bar D_{s}^{-}$ &   $1/6$ & $1/24$\\

 $D_{s}^{*+}\bar D_{s}^{*-}$ & $7/12$  &  $77/120$\\ 
\hline
\hline
\end{tabular*}
\end{table*}

As in Ref.~\cite{biz}, we take the form factor as follows
\begin{align}\label{formfac}
& & F(\bm q)^2=\frac{1+(R\,q_{\rm on})^2}{1+(R\,q)^2}, \ \  \ \ q \equiv |\bm q|, \ \ \ \
 q_{\rm on} = \frac{\lambda^{1/2}(s, m^2_{D_{i}}, m^2_{D_{i'}})}{2 \sqrt{s}}
\end{align}
where $R$ is a parameter. This factor is motivated by the Blatt-Weisskopf barrier penetration factor~\cite{blatt} which is often used in the parametrization of the width of resonances.
If the channel is closed, we take $q_{\rm on}=0$. Since $\tilde G_i(p^0)$ in Eq.~(\ref{Gtilda}) is logarithmically divergent, we have to make a subtraction and we do it such that the propagator has the pole at $M_{R}$. Hence the $D_{R}$ propagator rewritten as 
\begin{equation}\label{DRPpr}
D_{R}(p)=\dfrac{1}{p^{2}-M_{R}^{2} - \Pi'(p)}  
\end{equation}
with $\Pi'(p)=\Pi(p)-{\rm Re}~\Pi(M_{R})$. 

Next step we have to take into account the two resonances. We use $\sigma$ of Eq.~(\ref{crosssectionS}) to obtain the cross section for production of the channel $i~i'$, any of the $D^{0}\bar D^{0}$, $D^{+}\bar D^{-}$, $D^{0}\bar D^{*0}$, $D^{*0}\bar D^{0}$, $D^{*+}\bar D^{-}$, $D^{+}\bar D^{*-}$, $D^{*0}\bar D^{*0}$, $D^{*+}\bar D^{*-}$, $D_{s}^{+}\bar D_{s}^{-}$, $D_{s}^{+}\bar D_{s}^{*-}$, $D_{s}^{*+}\bar D_{s}^{-}$, $D_{s}^{*+}\bar D_{s}^{*-}$ channels. Then we must sum coherently the contribution of the two resonances  $\psi(4040)$  and $\psi(4160)$. Hence, $ f_{R} ~D_{R} ~A_{Ri}$ of Eq.~(\ref{crosssectionS}) gets substituted by $ f_{R1} ~D_{R1} ~A_{Ri} g_{1i}+  f_{R2}~ e^{i\phi}~D_{R2} ~A_{R2} g_{2i} $, which includes a relative phase factor between the two terms.
Then, the cross section is given by
\begin{equation}\label{crosssectionTWO}
\sigma_{i} =  \dfrac{1}{8 ~\pi} \frac{1}{s^{2}} \frac{1}{\sqrt{s}} e^{2}  (p')^{3}  \vert f_{R_{1}} ~D_{R_{1}} ~A_{1} g_{1i}+  f_{R_{2}}~ e^{i\phi}~D_{R_{2}} ~A_{2} g_{2i} \vert^{2}, 
\end{equation}
and vanishes for the channels which are closed at a given energy.

\section{Results}\label{result}

 The cross sections with the two resonances are obtained by means of Eq.~(\ref{crosssectionTWO}). As we mention in the section~\ref{formalism}, both the couplings  $e^+ e^-\to R_{1(2)}$ and $R_{1(2)}\to i~i'$ are different for particle 1 and 2, and then we have four parameters. The masses $M_{R_{1}}$ and $M_{R_{2}}$, enter into $D_{R_{1}}(p)$ and $D_{R_{2}}(p)$ in Eq.~(\ref{DRPpr}), are also changed a bit
with respect to the nominal ones, such as to obtain
the peak position as in the experiment. 
We will allow a relative phase between the two contributions, which means we will make the  $e^+ e^-\to R_{2}$ coupling complex. Hence $f_{R_{2}}$ will be complex to take into account the mixing of the two resonances as the experimental papers all do, that is  $f_{R_{2}} \to f_{R_{2}} e^{i\phi} $, thus we have the extra $f_{R_{1}}$, $f_{R_{2}}$ and  $\phi$ ($ \phi ~\epsilon~ [0,2 \pi] $) parameters. We have also $A_{1}$ and  $A_{2}$ real positive parameters. We shall also use the parameter $R$ in the form factor to control the convergence of the selfenergy integrals. So we have the $R_{1}$ and $R_{2}$ parameters for each resonance in Eq.~(\ref{formfac}). This means we have nine parameters in total. 
 
 \begin{figure}[h!]
  \centering
  \includegraphics[width=0.80\textwidth]{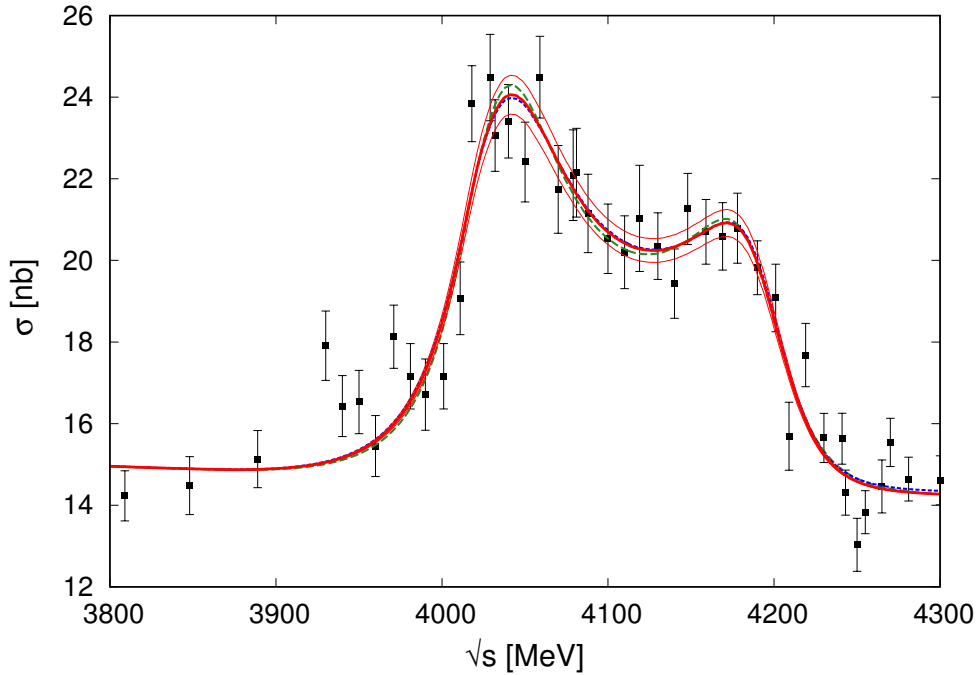}
  \caption{Total hadronic cross section of $\sigma (e^{+} e^{-} ~ \to ~hadrons)$ fitted to the experimental data from Fig.~1 in Ref. \cite{Mo:2010bw}. The solid red line corresponds the parameters Set I, dashed green line Set II and dotted blue line Set III. The most extreme thin lines provide the band of uncertainty of the fit evaluated as explained in the tex. }
  \label{figfit}
\end{figure}

\begin{figure}[h!]
  \centering
  \includegraphics[width=0.80\textwidth]{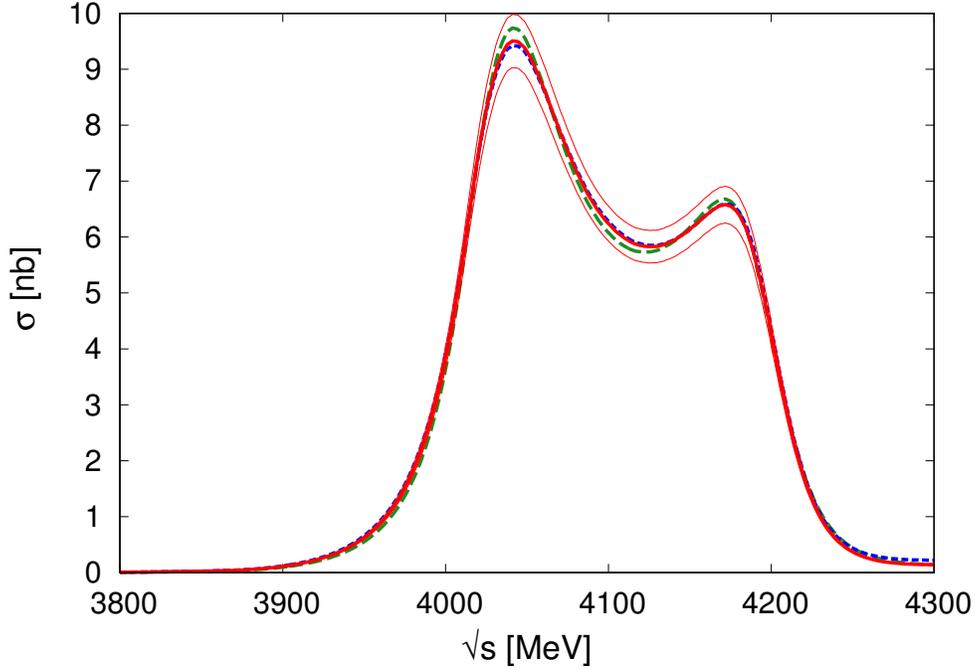}
  \caption{The total hadronic cross section of $\sigma (e^{+} e^{-} ~ \to ~hadrons)$ without background. The solid red line corresponds  the parameters Set I, dashed green line Set II and dotted blue line Set III. The most extreme thin lines provide the band of uncertainty of the fit evaluated as explained in the tex.}
  \label{figCSBsiz}
\end{figure}

 We  determine the nine parameters by performing a fit to the total hadronic cross section obtained as $\sigma (e^{+} e^{-} ~ \to ~hadrons)$ \cite{Mo:2010bw}. We get the experimental data from Fig.~1 of~\cite{Mo:2010bw} and include the same background as in Eq.~(3) of Ref. \cite{Mo:2010bw} with the same parameterization. We obtained three sets of values for the free parameters which are written in Table \ref{table1}, corresponding to different local minima of the $\chi^2$ in the multidimensional space. We also show there the errors in the parameters obtained in the fit. The total hadronic cross section including background for three different sets of parameter are shown in Fig. \ref{figfit}. As we see from the figure, the results of the total cross section of $\sigma (e^{+} e^{-} ~ \to ~hadrons)$ are similar with the three sets of parameters and they provide a good fit to the data, both above and below the peaks. In the figure, we also show the band of uncertainty in the fit, which is obtained making runs with random values for all the parameters within their uncertainty range, and then calculating the average and the dispersion. We also show the total  hadronic cross section without background in Fig.~\ref{figCSBsiz}, including the uncertainty band.
 
 \begin{table*}[h!]
\renewcommand\arraystretch{1.3}
\centering
\caption{\vadjust{\vspace{-2pt}}Three sets of fit data for nine parameters.}\label{table1}
\begin{tabular*}{0.85\textwidth}{@{\extracolsep{\fill}}|c|c|c|c|}
\hline
\hline
Parameters    & Set~I & Set~II & Set~III \\
\hline
 $M_{R_{1}}$ [MeV] & $4036.34 \pm 3.04$  & $4035.95 \pm 2.97 $ & $4036.73 \pm 3.21 $ \\
\hline
  $f_{R_{1}}$&$37.38 \pm 1.37 $ & $35.75 \pm 1.97 $&  $36.92 \pm 1.49 $ \\
\hline
 $ A_{1}^2 $    & $149.99 \pm 5.71 $ & $139.9 \pm 3.15 $ & $153.9 \pm 6.48$ \\
\hline
 $R_{1} $ [MeV$^{-1}$]  & $(8.71 \pm 4.36) \times 10^{-3} $ & $(1.19 \pm 0.59 ) \times 10^{-2}$ & $(7.41 \pm 3.71 )\times 10^{-3} $\\
\hline
 $M_{R_{2}}$ [MeV] & $4197.5\pm 7.95 $ & $4195.63 \pm 4.56 $ & $4199.17 \pm 11.21 $ \\
\hline
  $f_{R_{2}}$   & $30.92 \pm 3.31 $  & $27.94 \pm 11.76 $ & $25.47 \pm 8.19 $ \\
\hline 
   $ A_{2}^{2}$     &  $53.91 \pm 7.35 $  & $56.12 \pm 7.85 $ & $53.91 \pm 14.56$ \\
\hline
 $R_{2} $[MeV$^{-1}$] & $(2.68 \pm 1.34 ) \times 10^{-3} $ & $(2.23 \pm 1.12) \times 10^{-3}$& $(1.71 \pm 0.86) \times 10^{-3} $ \\
\hline 
  $\phi $ [radian]    & $3.408 \pm  0.283$   & $3.207 \pm 0.683 $ & $3.133 \pm 0.474 $ \\
\hline
\hline
\end{tabular*}
\end{table*}

\begin{figure}[h!]
  \centering
  \includegraphics[width=0.80\textwidth]{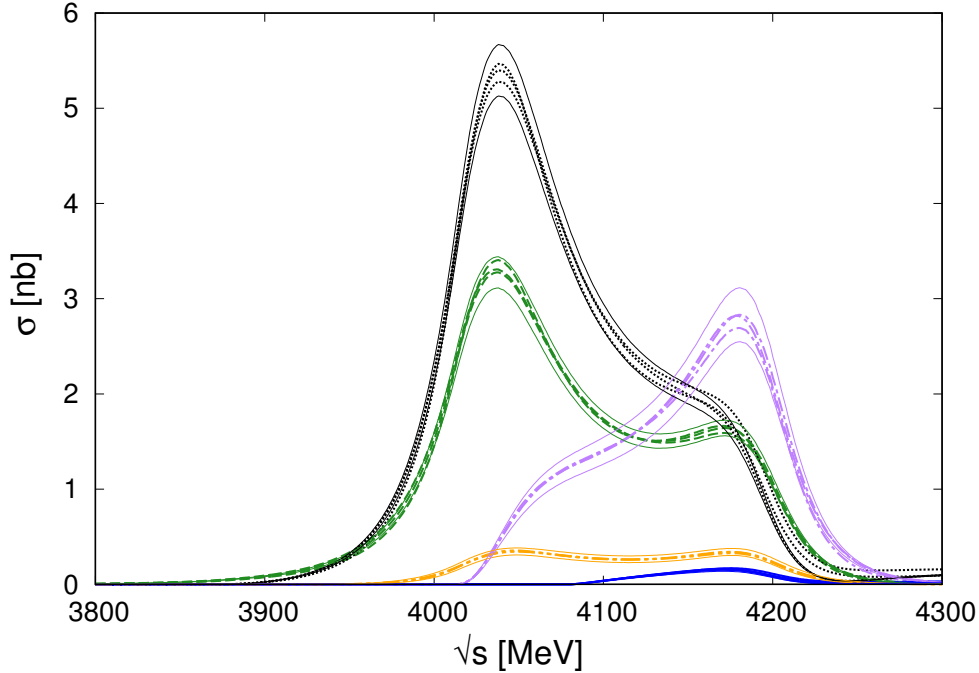}
  \caption{Different hadronic cross section of  $\sigma (e^{+} e^{-} ~ \to ~i~i')$ without bacground. The dashed green lines correspond to the $\sigma (D \bar{D})$, the black dotted lines  the $\sigma (D \bar{D}^{*})$, the purple dash-dotted lines the $\sigma (D^{*} \bar{D}^{*})$, the orange two dot-dashed lines the $\sigma (D_{s}^{+}\bar D_{s}^{-})$ and the blue solid lines the $\sigma (D_{s} \bar{D}_{s}^{*})$ for Set I, Set II and Set III. The most extreme thin lines provide the band of uncertainty of the fit evaluated as explained in the tex.}
  \label{figCSdif}
\end{figure}

 We also show the different contributions of the hadronic cross section in Fig.~\ref{figCSdif}. The $\sigma (D \bar{D})$ includes two contributions, as the sum of $D^{0}\bar D^{0}+D^{+}\bar D^{-}$, the $\sigma (D \bar{D}^{*})$ four contributions, as the sum of $D^{0}\bar D^{*0}+D^{*0}\bar D^{0}+D^{*+}\bar D^{-}+D^{+}\bar D^{*-}$, the $\sigma (D^{*} \bar{D}^{*})$ two contributions $D^{*0}\bar D^{*0}+D^{*+}\bar D^{*-}$, the $\sigma (D_{s} \bar{D}_{s}^{*})$ two contributions, as the sum of $D_{s}^{+}\bar D_{s}^{*-}+D_{s}^{*+}\bar D_{s}^{-}$  meson-meson states. The $\sigma (D_{s}^{+}\bar D_{s}^{-})$ includes only the $D_{s}^{+}\bar D_{s}^{-}$  meson-meson state. As we see in Fig.~\ref{figCSdif}, the results for different sets of parameters do not change much. Now let us look in detail at the figure. The biggest contributions come from the $\sigma (D \bar{D}^{*})$ and  the $\sigma (D \bar{D})$  for the  $\psi(4040)$. Conversely, the $\sigma (D^{*} \bar{D}^{*})$ gives the biggest contribution in the region of the $\psi(4160)$, while the contribution from the other channels is small in the whole region. We have not considered the $D^{*}_s \bar D^{*}_s$ production which only contributes a small quantity at the very end of the $\psi(4160)$ resonance tail.

We refrain from a detailed comparison with experimental data which are somewhat conflicting in the different experiments. Yet, it is instructive to look at the most recent measurements in Ref.~\cite{Zhukova:2017pen}. By looking at Fig.~16 of Ref.~\cite{Zhukova:2017pen} we can see that the $D^{+} D^{*-}$ channel peaks around 4040~MeV as in our Fig.~\ref{figCSdif} and has a shoulder around 4160~MeV, as we also find in Fig.~\ref{figCSdif}. These features are also observed in Fig.~4 of Ref.~\cite{Aubert:2009aq}. On the other hand, the $D^{*+} D^{*-}$ channel in Fig.~16 of Ref.~\cite{Zhukova:2017pen} peaks around 4150--4200~MeV, as is also the case of our contribution of that channel in Fig.~\ref{figCSdif}. The strength at the peaks of these two contributions filtering just the $D^+ D^{*-}$ and $D^{*+} D^{*-}$ channels in Fig.~\ref{figCSdif} is $ \displaystyle  \sim \frac{1}{4}  5.5  / \frac{1}{2}  3  = 0.92$  versus  $\sim 4/3 = 1.33$ in Ref.~\cite{Zhukova:2017pen}. The agreement is fair.

With the caveat on the problem of conflict in the experiment, we nevertheless compare in Table~\ref{table:new} our results with the experimental ones of Ref.~\cite{Bai:1999pk} and the theoretical ones, also using a $^3 P_0$ model, of Ref.~\cite{Barnes:2005pb}. Note that the comparison is not all fair since we obtain the results of Fig.~\ref{figCSdif} from the interference of the two resonances. Our results in Table~\ref{table:new} refer to values of Fig.~\ref{figCSdif} at the peak of the $\psi(4040)$ and $\psi(4160)$ including the interference. The individual contributions in the theory of Ref.~\cite{Barnes:2005pb} come from single particle states and in the experiment from an analysis of data in Ref.~\cite{Zhukova:2017pen}, which are also in conflict with those of Ref.~\cite{CroninHennessy:2008yi}. Hence, caution should be taken in the comparison to data in Table~\ref{table:new}.

\begin{table*}[htb!]
\renewcommand\arraystretch{1.2}
\centering
\caption{\vadjust{\vspace{-2pt}}
Ratios of branching fractions for the $\psi(4040)$ and $\psi(4160)$ resonances.}\label{table:new}
 \begin{tabular*}{1.0\textwidth}{@{\extracolsep{\fill}}cccc}
\hline
\hline
    & Ref.~\cite{Aubert:2009aq} & Ref.~\cite{Barnes:2005pb}   & Our results \\ 
\hline
 1) $\mathcal{B}(\psi(4040)\to D  \bar D)/\mathcal{B}(\psi(4040)\to D^* \bar D)$ \  &  0.24 $\pm$ 0.05 $\pm$ 0.12   & 0.003  & 0.54 $\pm$ 0.04\\ 
 2) $\mathcal{B}(\psi(4040)\to D^* \bar D^*)/\mathcal{B}(\psi(4040)\to D^* \bar D)$ \ &  0.18 $\pm$ 0.14 $\pm$ 0.03   & 1.0   & 0.18 $\pm$ 0.02\\ 
 3) $\mathcal{B}(\psi(4160)\to D  \bar D)/\mathcal{B}(\psi(4160) \to D^* \bar D^*)$ \ &  0.02 $\pm$ 0.03 $\pm$ 0.02  & 0.46  & 0.56 $\pm$ 0.06 \\ 
 4) $\mathcal{B}(\psi(4160)\to D^* \bar D)/\mathcal{B}(\psi(4160) \to D^* \bar D^*)$ \ &  0.34 $\pm$ 0.14 $\pm$ 0.05   & 0.011  & 0.60 $\pm$ 0.07\\ 
\hline
\hline
\end{tabular*}
\end{table*}

In addition to the caveat about comparing magnitudes which are not exactly the same, as discussed above, we should make some extra comments. The first one is that the theoretical errors quoted in table~\ref{table:new} are only statistical. There should be systematic errors tied to the assumptions done, but these are more difficult to quantize. Our assumptions relying on the $^3 P_0$ model to hadronize the $c \bar c$ state are rather standard and we do not rely upon a theoretical quark model but take input from experiment. Yet, there seems to be worse agreement in the case of the $\psi(4160)$ and part of it could be blamed on an approximation made in this case. Indeed, as shown in Eq.~(A.8) of Ref.~\cite{biz}, for the case of $d$-wave resonances there is the possibility to have internal angular momentum $l=1$ and $l=3$. The $l=3$ case involves quark matrix elements of $j_3 (\hat q)$ versus $j_1 (\hat q)$ for $l=1$, and also goes with $Y_{3 \mu}(\hat q)$ instead of $Y_{1 \mu} (\hat q)$ implicit in Eq.~(\ref{eq:19}). Making general assumptions of dominance of the lowest possible orbital angular momentum, the $l=3$ component is neglected in~\cite{biz} and here for the $d$-wave state. For the $s$-wave state only $l=1$ is possible and that approximation is not done. In this latter case the weights $g_{R, i}^2$ of table~\ref{gvalues} coincide with those of Refs.~\cite{DeRujula:1976gpz,vanBeveren:2009hb,vanBeveren:1979bd}. In any case, we think that a better comparison with data, given the interference of the two resonances, should be done with the distributions for the different $D^{(*)} \bar D^{(*)}$,$D_s^{(*)} \bar D_s^{(*)}$, channels as a function of the energy, in the line discussed above comparing with data of Refs.~\cite{Aubert:2009aq,Zhukova:2017pen}. New and more precise data improving on present ones and solving actual discrepancies would help to clarify the issue.

There are discrepancies of our results with the experimental analysis, particularly in the $D \bar D / D^* \bar D^*$ ratio of the $\psi(4160)$ where both the results of~\cite{Barnes:2005pb} and the present ones strongly disagree with the experimental analysis. In the other cases the present results are closer to the analysis than those of Ref.~\cite{Barnes:2005pb}.

The formalism used in the present work is similar to the one used in~\cite{Piotrowska:2018rzl} in the study of the $\psi(4040)$ resonance, earlier used also in the study of the lineshape of the $\psi(3770)$~\cite{Coito:2017ppc}, renormalizing the vector propagators by the meson-meson selfenergy. Apart from some technical details, the major difference is that the couplings of the resonance to the $D^{(*)}\bar D^{(*)}$ channels are taken from experiment (actually from the ratios of Table~\ref{table:new}), while we have determined them theoretically by means of the $^3 P_0$ model. A second major difference is that we have studied the $\psi(4040)$ in connection with the $\psi(4160)$ due to the strong interference between the two, while only the $\psi(4040)$ is considered in~\cite{Piotrowska:2018rzl}, although it is mentioned that for the purpose of their work, the investigation of a possible associated resonance, the consideration of the $\psi(4160)$ does not change much the results.

We can also do the compositeness test. In order to calculate the strength of the original vector when it is dressed by the meson-meson components we use the following formula (See detailed analysis in Ref.~\cite{biz})	
\begin{align}\label{Z1}
Z&=\frac{1}{1-\frac{\partial Re\Pi(p^2)}{\partial p^2}\Big|_{p^2=M^2_R}}.
\end{align}
On the other hand $1-Z$ provides the meson-meson strength of the dressed vector.\footnote{In Ref.~\cite{biz} a thorough discussion is made of the meaning of the strengths of the meson-meson channel, which measures the weight of each channel in the wave function, but cannot be associated to a probability, except if the channel is closed. }
If $\frac{\partial Re\Pi(p^2)}{\partial p^2}$ is quite smaller than $1$, we can make the following expansion 
\begin{align}
1-Z \simeq -\frac{\partial Re\Pi}{\partial p^2}\Big|_{p^2=M^2_R},
\end{align}	 
where $-\frac{\partial Re\Pi}{\partial p^2}\Big|_{p^2=M^2_R}$ reads as the meson-meson strength in the wave function and it can be written for each channel as
\begin{align}\label{proba2}
P_{(MM)i}\simeq-\frac{\partial Re\Pi_i(p^2)}{\partial p^2}\Big|_{p^2=M^2_R},
\end{align}
where $\Pi_i$ is the contribution of the $i$-th channel to $\Pi$. We calculate the meson-meson weight and show the results using the results of Set I in Table~\ref{set1P40} and Table~\ref{set1P60} for  the $\psi(4040)$ and $\psi(4160)$, respectively.
The value of $Z$ shown there is obtained from Eq.~(\ref{Z1}) and includes the contributions from all channels. 

\begin{table*}[h!]
\renewcommand\arraystretch{1.2}
\centering
\caption{\vadjust{\vspace{-2pt}}Meson-meson probabilities in the $\psi(4040)$ wave function for Set I.}\label{set1P40}
\begin{tabular*}{0.70\textwidth}{@{\extracolsep{\fill}}cccc}
\hline
\hline
   Channels                  & $-\frac{\partial\Pi}{\partial p^2}\big|_{p^2=M^2_{\psi(4040)}}$ &$P_{(MM)}$          & $Z$\\
                               \hline
    $D^0\bar D^0$            &$(-1.499+1.065i)\times 10^{-2}$  &$-1.499 \times 10^{-2}$                    & \\
    \hline
    $D^+D^-$                 &$(-1.434+1.054i)\times 10^{-2}$  &$-1.434  \times 10^{-2}$                 & \\
    \hline
    $D^0\bar D^{\ast0}+c.c$  &$(-2.816+3.499i)\times 10^{-2}$  &$-2.816  \times 10^{-2}$                     & \\
    \hline
    $D^+\bar D^{\ast-}+c.c$  &$(-2.654+3.437 i)\times 10^{-2}$  &$-2.654 \times 10^{-2}$                     & \\
    \hline
    $D^{\ast0}\bar D^{\ast0}$&$(-1.572 \times 10^{-3}+ 3.267 i\times 10^{-2})$  &$-1.572 \times 10^{-3}$      & \\
    \hline
    $D^{\ast+}D^{\ast-}$     &$(2.108 \times 10^{-3}+ 2.968 i \times 10^{-2})$  &$2.108 \times 10^{-3}$         & \\
    \hline
    $D^+_sD^-_s$             &$(-4.026+7.377i)\times 10^{-3}$  &$-4.026 \times 10^{-3}$                     & \\
    \hline
    $D_s^+D_s^{\ast-}+c.c$   &$(1.591 \times 10^{-3}+ 2.366i \times 10^{-5})$  &$1.591 \times 10^{-3}$       & \\
    \hline
    $D^{\ast+}_sD^{\ast-}_s$ &$(8.467 \times 10^{-4}+ 4.427i \times 10^{-6})$  &$8.467 \times 10^{-4}$       & \\
    \hline
    Total                    &$(-0.0851-0.0160i)$  &$-0.0851$                    &$0.91$\\
\hline
\hline
\end{tabular*}
\end{table*}

\begin{table*}[h!]
\renewcommand\arraystretch{1.2}
\centering
\caption{\vadjust{\vspace{-2pt}}Meson-meson probabilities in the $\psi(4160)$ wave function for Set I.}\label{set1P60}
\begin{tabular*}{0.70\textwidth}{@{\extracolsep{\fill}}cccc}
\hline
\hline
   Channels                  & $-\frac{\partial\Pi}{\partial p^2}\big|_{p^2=M^2_{\psi(4160)}}$ &$P_{(MM)}$          & $Z$\\
                               \hline
    $D^0\bar D^0$            &$(-1.155\times 10^{-2}+5.296i\times 10^{-3})$  &$-1.155 \times 10^{-2}$                    & \\
    \hline
    $D^+D^-$                 &$(-1.107\times 10^{-2}+5.339i \times 10^{-2})$  &$-1.107  \times 10^{-2}$                 & \\
    \hline
    $D^0\bar D^{\ast0}+c.c$  &$(-6.171+5.157i)\times 10^{-2}$  &$-6.171  \times 10^{-2}$                     & \\
    \hline
    $D^+\bar D^{\ast-}+c.c$  &$(-5.934+5.141 i)\times 10^{-3}$  &$-5.934\times 10^{-3}$                     & \\
    \hline    
    $D^{\ast0}\bar D^{\ast0}$&$(-2.057 +3.648i ) \times 10^{-2}$  &$-2.057 \times 10^{-2}$      & \\
    \hline    
    $D^{\ast+}D^{\ast-}$     &$(-1.939 + 3.624i ) \times 10^{-2}$  &$-1.939 \times 10^{-2}$         & \\
    \hline    
    $D^+_sD^-_s$             &$(-4.471+5.007i)\times 10^{-3}$  &$-4.471 \times 10^{-3}$             & \\
    \hline    
    $D_s^+D_s^{\ast-}+c.c$   &$(-1.127+4.361i ) \times 10^{-3}$  &$-1.127 \times 10^{-3}$       & \\
    \hline
    $D^{\ast+}_sD^{\ast-}_s$ &$(8.845 \times 10^{-3}+ 1.423i \times 10^{-4})$  &$8.845 \times 10^{-3}$       & \\
    \hline
    Total                    &$(-7.145 \times 10^{-2}+0.103i)$  &$-7.145\times 10^{-2}$                    &$0.93$\\
\hline
\hline
\end{tabular*}
\end{table*}

As we can see in Tables~\ref{set1P40} and~\ref{set1P60}, we obtain some negative results for the $P_{(MM)}$, which should not come as a big surprise since the weights have not to be identified with a probability, as discussed in detail in~\cite{biz}. For the closed channels, where the identification of weight with probability is fair, we obtain positive numbers. The interesting result, however, is that all these numbers are very small, which make these vector mesons to qualify as mostly $c \bar c$ states with a very small meson-meson cloud component.

\section{Summary and conclusions}\label{conclusion}
We have made a fit to the $e^+ e^- \to hadron$ data in the region of the $\psi(4040)$ and $\psi(4160)$ resonances, considering the hadronic channels $D \bar D$, $D \bar D^*$, $D^* \bar D$, $D^* \bar D^*$, $D_s \bar D_s$, $D_s \bar D^*_s$, $D^*_s \bar D_s$, $D^*_s \bar D^*_s$. 
We have taken into account the renormalization of the vector mesons by including the meson-meson selfenergy, and the scheme provides automatically the cross sections into the different channels. We had some freedom fitting the data by means of a few parameters, but the relative weight of the different channels was calculated theoretically by means of the $^3P_0$ model, hence the relative ratios for production of the different channels is a prediction of the theory. The sometimes conflicting experimental results do not allow a quantitative comparison with experiment, but the agreement with the most recent experimental data is fair. We have also explored a side effect of the theory by evaluating the weights of the different meson-meson components in the wave function and determined that the $\psi(4040)$ and $\psi(4160)$ have very small meson-meson components and largely qualify as $c \bar c$ states. Improved rates of production of these channels in the future should allow for a more quantitative comparison, and agreement or disagreement with data can shed light on possible mixing of the $c \bar c$ charmonium states with other configurations, a topic of current interest.

\section{ACKNOWLEDGEMENT}
This work is partly supported by the Spanish Ministerio de Economia y Competitividad and European FEDER funds under Contracts No. FIS2017-84038-C2-1-P B and No. FIS2017-84038-C2-2-P B, and the
Generalitat Valenciana in the program Prometeo II-2014/068, and the project Severo Ochoa
of IFIC, SEV-2014-0398.
The work of N. I. was partly supported by JSPS Overseas Research Fellowships and JSPS KAKENHI Grant Number JP19K14709.


\end{document}